\documentclass[%
 reprint,
 superscriptaddress,
 amsmath,amssymb,
 aps,
]{revtex4-1}

\usepackage{graphicx}
\usepackage{dcolumn}
\usepackage{bm}

\begin{document}

\preprint{APS/123-QED}

\title{Quantum interference and the time-dependent radiation of nanojunctions}

\author{Michael Ridley}
\affiliation{%
School of Physics and Astronomy, Tel-Aviv University, Tel-Aviv 69978, Israel \looseness=-1}%

\author{Lev Kantorovich}
\affiliation{
Department of Physics, King’s College London, Strand, London, WC2R 2LS, United Kingdom \looseness=-1}%

\author{Robert van Leeuwen}
\affiliation{Department of Physics, Nanoscience Center, University of Jyv{\"a}skyl{\"a}, 40014 Jyv{\"a}skyl{\"a}, Finland \looseness=-1}

\author{Riku Tuovinen}
\affiliation{
QTF Centre of Excellence, Turku Centre for Quantum Physics, Department of Physics and Astronomy, University of Turku, 20014 Turku, Finland}
\affiliation{QTF Centre of Excellence, Department of Physics, P.O. Box 43, 00014 University of Helsinki, Finland \looseness=-1}
%


\begin{abstract}
Using the recently developed time-dependent Landauer-B{\"u}ttiker formalism and Jefimenko's retarded solutions to the Maxwell equations, we show how to compute the time-dependent electromagnetic field produced by the charge and current densities in nanojunctions out of equilibrium. We then apply this formalism to a benzene ring junction, and show that geometry-dependent quantum interference effects can be used to control the magnetic field in the vicinity of the molecule. Then, treating the molecular junction as a quantum emitter, we demonstrate clear signatures of the local molecular geometry in the non-local radiated power.

\end{abstract}

\maketitle


\section{\label{sec:intro}Introduction}

Quantum transport properties of molecular structures have multiple applications in modern information-processing technologies \cite{xiang_molecular-scale_2016, thoss_perspective_2018, gehring_single-molecule_2019}. Molecular junctions also provide important insights into a wide range of physical effects in nonequilibrium many-body systems at very short timescales. In these systems, the steady state response to externally applied fields can encode information about quantum interference effects \cite{solomon_understanding_2008, hansen_interfering_2009, solomon_exploring_2010, markussen_relation_2010, huang_controlling_2018}, electron-electron interactions~\cite{Thygesen2008, Ness2011, Myohanen2012, krivenko_dynamics_2019-2, Hopjan2018, Talarico2020, tuovinen_comparing_2020, Cosco2020} and current fluctuations \cite{blanter_shot_2000, moskalets_heat_2009}. However, THz intramolecular transport processes are increasingly relevant for determining the operational frequencies of nanodevices beyond the steady state, which may be related, for instance, to dynamical symmetries in periodically driven structures~\cite{Foieri2010, Ludovico2012, schaller_single-electron_2013, ridley_time-dependent_2017}, spin-flip processes \cite{frustaglia_aharonov-bohm_2004, perfetto_spin-flip_2008}, transport statistics~\cite{tang_full-counting_2014, tang_full-counting_2017, ridley_numerically_2018-1, ridley_numerically_2019}, and electron traversal times~\cite{fevrier_tunneling_2018, ridley_electron_2019}. In addition, the optical properties of irradiated molecular structures have wide-ranging uses resulting from their photoluminescence, photodetection and frequency conversion potential \cite{galperin_optical_2006,bonaccorso_graphene_2010,galperin_photonics_2017}.

Most studies on magnetic effects in nanojunctions have focused on electron transport induced by external magnetic fields \cite{kim_tuning_2010,rai_magnetic_2012,hayakawa_large_2016,maslyuk_enhanced_2018}, with longstanding interest in the Aharonov-Bohm effect and related phenomena \cite{aharonov_significance_1959,bachtold_aharonovbohm_1999,duca_aharonov-bohm_2015}. However, there has recently been a growing interest in electromagnetic fields induced by the electronic currents and charge densities in the molecule \cite{tagami_electronic_2003,tsuji_large_2007,maiti_externally_2015,patra_modulation_2017,patra_bias-induced_2019,Zhang2020}. Experiments have been performed showing laser-induced circular currents and associated magnetic fields in the $1$~mT regime \cite{pershin_laser-controlled_2005,rasanen_optimal_2007,eckart_ultrafast_2018}, and much higher field strengths ($\sim 1$~T) appear to be possible in nanosolenoids \cite{zhang_nano-solenoid_2013,xu_riemann_2016,dyachkov_gold_2019}. This makes a study of molecular current-induced magnetism timely and useful. We note the recent studies by \citet{Zhang2020} of the steady-state angular momentum radiation of molecular junctions which relate this non-local observable to localized intramolecular transitions \cite{zhang_quantum_2020,zhang_far-field_2020}. However, a fully time-dependent framework for describing the radiative response to the current in such nanodevices is still lacking. 

In principle, a fully self-consistent and microscopic coupling of light and matter is needed to capture the interplay of quantum dynamics and local electromagnetic fields. Recently, this has been considered in the context of quantum-electrodynamical density-functional theory~\cite{Ruggenthaler2014, Ruggenthaler2018, Jestaedt2019, Haugland2020b}, coupled-cluster theory~\cite{Haugland2020a, Haugland2020b}, or the nonequilibrium Green's function (NEGF) approaches~\cite{Melo2016, Sentef2018, Zhang2020}. However, in many cases of interest to the experimentalist, ballistic transport dominates, making a noninteracting tight-binding approach sufficient for a description of quantum conductance~\cite{agrait_quantum_2003,heedt_ballistic_2016,annadi_quantized_2018,aprojanz_ballistic_2018}. The Landauer-B{\"u}ttiker (LB) formalism has provided an accurate theoretical description of the ballistic transport response to static biases in the steady state regime. 

Recently, a time-dependent extension of the LB approach based on NEGF has been developed incorporating transient effects resulting from the switch-on of a bias, which may be an arbitrary function of time \cite{tuovinen_time-dependent_2013,tuovinen_time-dependent_2014,ridley_current_2015}. This time-dependent Landauer-B{\"u}ttiker (TD-LB) method has been applied to the study of superconductivity \cite{tuovinen_time-dependent_2016,tuovinen_distinguishing_2019}, impurity models \cite{Rocha2015, tuovinen_time-resolved_2019}, double quantum dots~\cite{fukadai_transient_2018}, nanowires~\cite{ridley_calculation_2016, Fukadai2019}, energy currents~\cite{Eich2016, tuovinen_phononic_2016, covito_transient_2018}, systems with spatial and dynamical disorder~\cite{ridley_fluctuating-bias_2016,ridley_electron_2019}, time-dependent quantum noise and electron traversal times~\cite{ridley_partition-free_2017, ridley_electron_2019} and periodically-driven molecular junctions~\cite{ridley_time-dependent_2017}. In all of these studies, the dynamical response of charge and current densities to external fields was computed. These generate local time-dependent electromagnetic fields, with a small retardation in the response time due to relativistic causality. Thus, the static Coulomb and Biot-Savart laws for the $\textbf{E}$ and $\textbf{B}$ fields must be replaced with causal expressions giving these fields in terms of their sources. The correct formulas expressing the time-dependent electromagnetic field components in terms of their sources was published by Jefimenko in 1966 \cite{jefimenko_electricity_1966}.

In this work, we use the Jefimenko formulas in combination with the TD-LB formalism to compute the time-dependent electromagnetic fields in the vicinity of benzene molecules coupled to metal electrodes within a tight-binding approach. In Section~\ref{TDLB_Formalism_Sec}, we describe the TD-LB method used for the calculation of Green's functions and electrode currents resulting from the switch-on of a bias. Section \ref{TDB_Formalism_Sec} contains a derivation of the exact formal expression for the magnetic field generated in a junction; this is then shown to reduce to an expression in terms of interface currents in the electrodes and individual bond currents resulting from internal electron transfer in the molecule, and then we show how to use these as source terms for the local fields. Then, in Section \ref{TD_plots_Sec} we apply this formalism to a molecular junction composed of a benzene molecule coupled to the electrodes in the para, meta and ortho configurations. We find quantitative and qualitative differences between the time-dependent electronic behaviour of all three types of junction, which become enhanced in the case of strong biases. In addition, we show the detailed temporal relaxation of the magnetic field to a steady state value corresponding to the constant applied bias in the vicinity of the benzene ring. We investigate the steady-state transport properties of the system in Section \ref{sec:interference}. There we show the step-like current-voltage characteristics for both the electrode currents and ring currents, and we also plot the maximum magnetic field as a function of voltage. In these plots we see novel resonances in the ortho and meta cases, which arise from quantum interference between electron pathways through the asymmetrically-coupled benzene ring. Finally, Section \ref{sec:radiation} includes a calculation of the Poynting vector for the rate of flow of electromagnetic energy density out of the molecular region. We then map the detailed angular distribution of radiated energy out of the molecule. We find a novel dependence of the radiation flux profile on quantum interference effects caused by the local molecular geometry, and mediated by variations in the local magnetic fields.

\section{Formalism} \label{Formalism_Sec}

\subsection{The TD-LB Formalism}\label{TDLB_Formalism_Sec}

We consider transport in a generic lead-molecule-lead junction driven out of equilibrium by the switch-on of a bias at the quench time $t_{0}$, described by a noninteracting Hamiltonian \cite{tuovinen_time-dependent_2013,ridley_formal_2018}:
\begin{align}\label{Hamiltonian}
\hat{H}\left(z\right) & =\underset{k\alpha\sigma}{\sum}\varepsilon_{k\alpha}\left(z\right)\hat{d}_{k\alpha\sigma}^{\dagger}\hat{d}_{k\alpha\sigma}+\underset{mn\sigma}{\sum}T_{mn}\left(z\right)\hat{d}_{m\sigma}^{\dagger}\hat{d}_{n\sigma} \nonumber \\
& + \underset{m,k\alpha\sigma}{\sum}\left[T_{m,k\alpha}\left(z\right)\hat{d}_{m\sigma}^{\dagger}\hat{d}_{k\alpha\sigma}+T_{k\alpha,m}\left(z\right)\hat{d}_{k\alpha\sigma}^{\dagger}\hat{d}_{m\sigma}\right] .
\end{align}
The first term of Eq. (\ref{Hamiltonian}) describes the electronic energy states of the leads, $k\alpha$. In this paper, the leads are denoted by $\alpha$, which may take on the values $S$ or $D$ to denote the source or drain, respectively. The second term of the Hamiltonian describes the hopping between sites internal to the molecular region and the third term describes the coupling between the molecule and the leads. The argument $z$ refers to times defined on the Konstantinov-Perel' time contour $\gamma$ \cite{konstantinov_graphical_1960,keldysh_diagram_1964,stefanucci_nonequilibrium_2013,kantorovich_generalized_2020}, composed of an upper branch $C_{-}$ running in the direction of increasing time from $t_{0}$ to $t$, then along a lower branch $C_{+}$ which runs backwards from $t$ to $t_{0}$. The equilibration of the initial state is represented by the vertical imaginary time branch $C_{M}$ which runs from $t_{0}$ to $t_{0}-i\beta$, where $\beta\equiv1/k_{B}T$ is the inverse temperature (we use units in which $\hbar=1$). 

We work in the partition-free quench framework, which means that the lead-molecule coupling terms $T_{k\alpha,m}\left(z\right)$ are nonzero for all values of $z \in C$, i.e. the molecule and the leads are coupled during equilibration ($t<t_{0}$) as well as during the transport ($t\ge t_{0}$). For the purposes of this paper, we also assume no contour-time dependence in these couplings, i.e. $T_{k\alpha,m}\left(z\right) \equiv T_{k\alpha,m}$. We also drop any time-dependence in the intramolecular hopping integrals, $T_{mn}\left(z\right) \equiv T_{mn}$, although we have previously considered time-dependent molecular energies within the TD-LB formalism in Refs. \cite{tuovinen_time-dependent_2014,ridley_fluctuating-bias_2016}.

\begin{figure}[t]
    \centering
    \includegraphics[width=0.45\textwidth]{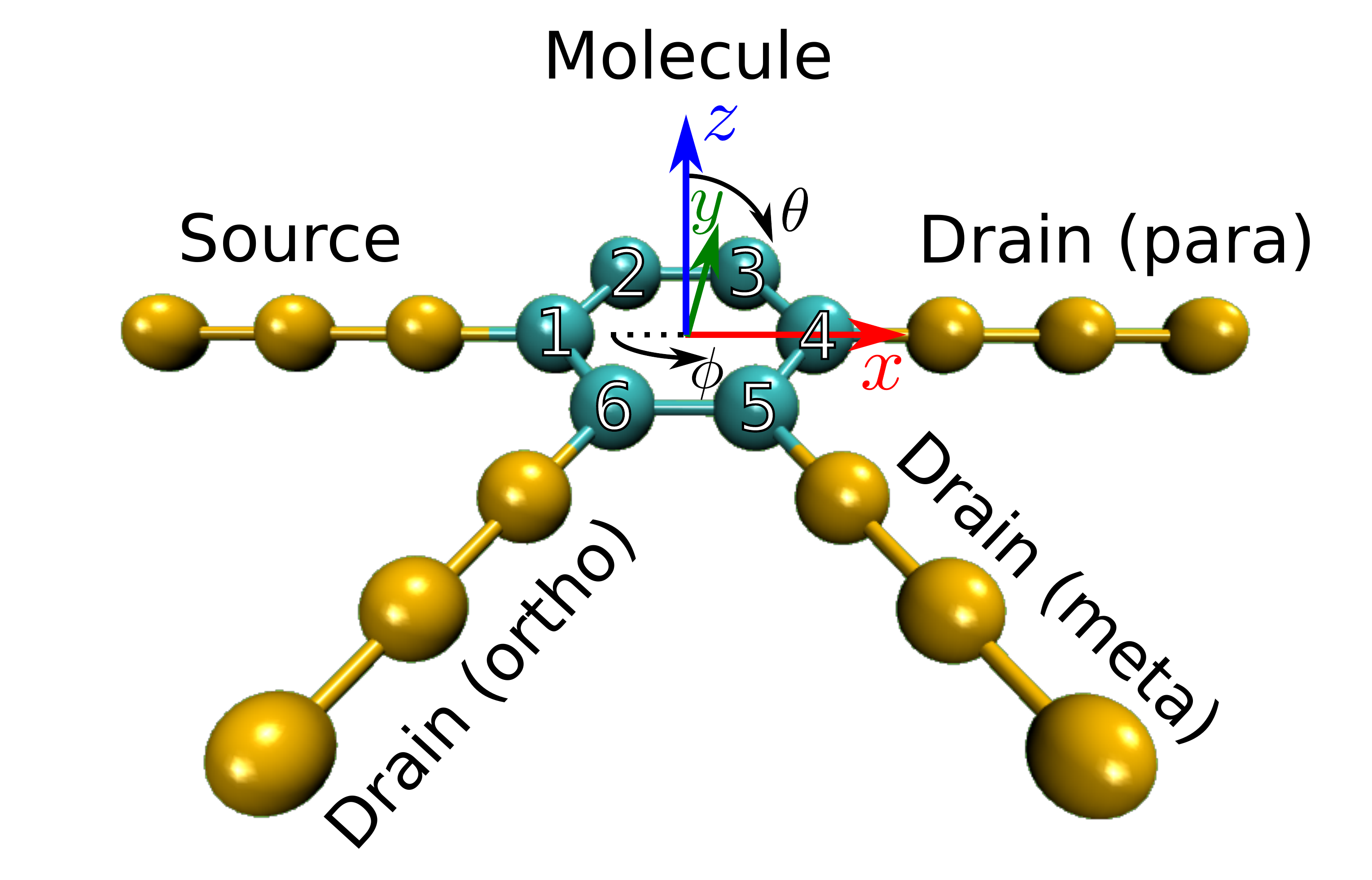}
    \caption{Schematic of the molecular junction where the central region is a benzene molecule connected to semi-infinite metallic electrodes. We consider three cases of source and drain electrodes being attached in the \emph{para}, \emph{meta}, or \emph{ortho} configuration. The atomic sites within the molecule are labeled from $1$ to $6$. Transport direction in the para case is assigned with the $x$ coordinate and the molecule lies in the $xy$-plane. The corresponding polar ($\theta$) and azimuthal ($\phi$) angles of the underlying three-dimensional coordinate system are considered in Sec.~\ref{sec:radiation}.}
    \label{fig:schematic}
\end{figure}

An example of a molecular junction described by Eq. (\ref{Hamiltonian}) is shown in Fig.~\ref{fig:schematic}, where the central molecular region is a six-site benzene ring. We give a suitable parametrization for this type of molecule in Sec.~\ref{sec:results}, accompanied by numerical simulation results. Such configurations can be realized experimentally in mechanically-controlled break junctions \cite{kaneko_fabrication_2010,arroyo_signatures_2013} or in lithographically etched transistors \cite{jia_quantum_2018,gee_nanoscale_2020}. 

To model the bias switch-on process, we add a spatially homogeneous time-dependent shift to the lead energies on the horizontal contour branches at time $t_{0}$, which mimics the switch-on of a time-dependent bias in the leads \cite{jauho_time-dependent_1994,ridley_current_2015}: 
\begin{equation}\label{TD_energies}
\varepsilon_{k\alpha}\left(z\in C_{\mp}\right)	=	\varepsilon_{k\alpha}+V_{\alpha}\left(t\right) .
\end{equation}
On the vertical contour branch the system is propagated by the Matsubara hamiltonian $\hat{H}^{M}$, which is equivalent to Eq.~(\ref{Hamiltonian}) with shifted lead energies $\varepsilon_{k\alpha}\left(z\in C_{M}\right)=\varepsilon_{k\alpha}-\mu_{\alpha}$, where $\mu_{\alpha}$ is the chemical potential of lead $\alpha$.

The crucial object in the NEGF formalism is the Green's function on the contour:
\begin{equation}\label{GF_def}
G\left(z_{1},z_{2}\right)_{ij}=-i\frac{\textrm{Tr}\left[e^{-\beta\hat{H}^{M}}\hat{T}_{\gamma}\left[\hat{d}_{i,H}\left(z_{1}\right)\hat{d}_{j,H}^{\dagger}\left(z_{2}\right)\right]\right]}{\textrm{Tr}\left[e^{-\beta\hat{H}^{M}}\right]}.
\end{equation}
In this expression, the two times $z_{1}$, $z_{2}$ may be located anywhere on $\gamma$, and the operator $\hat{T}_{\gamma}$ orders operator-valued functions of contour time with the latest on $\gamma$ to the left. The Green's function may then be projected onto the central (molecular) region (denoted $CC$) to obtain the matrix-valued function $\mathbf{G}_{CC}$, which satisfies the Kadanoff-Baym integro-differential equations of motion \cite{stefanucci_nonequilibrium_2013} with integral kernel given by the embedding self energy
\begin{equation}\label{Embedding_SE}
\left[\mathbf{\Sigma}_{CC}\left(z_{1},z_{2}\right)\right]_{mn}=\underset{k \alpha}{\sum}T_{m,k\alpha}\left[\mathbf{g}_{\alpha\alpha}\left(z_{1},z_{2}\right)\right]_{kk}T_{k\alpha,n} ,
\end{equation}
where $\mathbf{g}_{\alpha\alpha}$ is the Green's function of the decoupled lead $\alpha$. We now assume that the leads satisfy the wide-band limit approximation (WBLA), i.e. we neglect the energy dependence of the lead-molecule coupling. This assumption enables us to write down all components of the effective embedding self-energy in terms of the level-width matrix $\Gamma_{\alpha}$, defined as \cite{stefanucci_nonequilibrium_2013,tuovinen_time-dependent_2014,ridley_current_2015}:
\begin{equation}\label{Gamma_matrix}
\Gamma_{\alpha,mn}=2\pi\sum_{k}T_{m,k\alpha}T_{k\alpha,n}\delta\left(\varepsilon_{\alpha}^{F}-\varepsilon_{k\alpha}\right) ,
\end{equation}
where $\varepsilon_{\alpha}^{F}$ is the equilibrium Fermi energy of lead $\alpha$. 
Within the WBLA, these equations are linearized in terms of the effective Hamiltonian of the central region,
\begin{equation}\label{Effective_Hamiltonian}
\mathbf{h}_{CC}^{eff}\equiv\mathbf{h}_{CC}-\frac{i}{2}\underset{\alpha}{\sum}  \Gamma_{\alpha} .
\end{equation}
 The detailed derivation of the Green's function and self-energy components was published in Ref.~\cite{ridley_partition-free_2017}, and includes the following compact formula for the greater and lesser Green's functions:
\begin{widetext}
\begin{equation}\label{G_Lesser_Greater}
\mathbf{G}_{CC}^{\gtrless}\left(t_{1},t_{2}\right)=\mp i\int\frac{d\omega}{2\pi}f\left(\mp\left(\omega-\mu\right)\right)\underset{\beta}{\sum}\mathbf{S}_{\beta}\left(t_{1},t_{0};\omega\right)\Gamma_{\beta}\mathbf{S}_{\beta}^{\dagger}\left(t_{2},t_{0};\omega\right) ,
\end{equation}
where the upper (lower) signs on the right hand side correspond to the greater (lesser) components and we have introduced the matrix
\begin{align}\label{S_alpha}
& \mathbf{S}_{\alpha}\left(t,t_{0};\omega\right)\equiv e^{-i\mathbf{h}_{CC}^{eff}\left(t-t_{0}\right)}\left[\mathbf{G}_{CC}^{r}\left(\omega\right) -i\intop_{t_{0}}^{t}d\bar{t}e^{-i\left(\omega\mathbf{1}-\mathbf{h}_{CC}^{eff}\right)\left(\bar{t}-t_{0}\right)}e^{-i\psi_{\alpha}\left(\bar{t},t_{0}\right)}\right]
\end{align}
defined in terms of the retarded Green's function $\mathbf{G}_{CC}^{r}\left(\omega\right)=\left(\omega\mathbf{I}-\mathbf{h}_{CC}^{eff}\right)^{-1}$. The time-dependent voltage in the leads is contained in phase factors of the form:
\begin{equation}\label{phase}
\psi_{\alpha}\left(t_{1},t_{2}\right)\equiv\underset{t_{2}}{\overset{t_{1}}{\int}}d\tau\,V_{\alpha}\left(\tau\right) .
\end{equation} 

All components of the Green's function (corresponding to different combinations of pairs of contour branch times) can be calculated exactly in the two-time plane \cite{ridley_current_2015,ridley_fluctuating-bias_2016}. The quantum statistical expectation value of the current operator, setting the electronic charge $q=-1$, may be expressed in terms of the $\mathbf{S}_{\alpha}$, as~\cite{ridley_partition-free_2017}:
\begin{align}\label{Current_alpha}
I_{\alpha}\left(t\right) & = \frac{1}{\pi}\int d\omega f\left(\omega-\mu\right)\,\mbox{Tr}_{C}\left[2\mbox{Re}\left[i\Gamma_{\alpha}e^{i\omega\left(t-t_{0}\right)}e^{i\psi_{\alpha}\left(t,t_{0}\right)}\mathbf{S}_{\alpha}\left(t,t_{0};\omega\right)\right]-\Gamma_{\alpha}\underset{\beta}{\sum}\mathbf{S}_{\beta}\left(t,t_{0};\omega\right)\Gamma_{\beta}\mathbf{S}_{\beta}^{\dagger}\left(t,t_{0};\omega\right)\right] .
\end{align}
\end{widetext}

Eq.~\eqref{Current_alpha} is a closed expression for the time-dependent current within the WBLA, at the interface of the molecular device and the lead labelled by$\alpha$. This expression reduces to the traditional Landauer-B{\"u}ttiker formula in the case of a static bias in the long-time limit $t\to\infty$~\cite{ridley_current_2015}.

\subsection{Time-Dependent Electromagnetic Field Components}\label{TDB_Formalism_Sec}

We first wish to compute time-dependent magnetic field $\mathbf{B}\left(\mathbf{r},t\right)$ defined at an arbitrary spatial position $\mathbf{r}$ and time $t$. We note that the Biot-Savart law for the static magnetic field due to a steady current density $I\left(\mathbf{l}\right)$ flowing along a path $P$ (described by the set of points $\mathbf{l}\in P$) is given by:
\begin{equation}\label{Biot_Savart_static}
\mathbf{B}\left(\mathbf{r}\right)=\frac{\mu_{0}}{4\pi}\int_{P}\frac{I\left(\mathbf{l}\right)d\mathbf{l}\times\left(\mathbf{r}-\mathbf{l}\right)}{\left|\mathbf{r}-\mathbf{l}\right|^{3}} ,
\end{equation}
where $\mu_0$ is the vacuum permeability and $d\mathbf{l}$ is an element of the path taken by the current. In most textbook discussions of the relationship between the time dependent components of the electromagnetic field and their sources, the magnetic field is related to the time-varying electric field, via the displacement current term in the Amp\`{e}re-Maxwell law:
\begin{equation}\label{Ampere_Maxwell}
\nabla\times\mathbf{B}=\mu_{0}\mathbf{J}+\mu_{0}\epsilon_{0}\frac{\partial\mathbf{E}}{\partial t} .
\end{equation}
Although correct, this expression should not be used to derive the time-dependent generalization of Eq.~(\ref{Biot_Savart_static}), even though it can be combined with Helmholtz's theorem to give a spatial integral formula for the magnetic field \cite{griffiths_timedependent_1991}. This is because the electric field is not the localized physical source of magnetic field. Instead, one should use the formulation of Jefimenko, where the magnetic field is obtained from the relation $\mathbf{B}=\nabla\times\mathbf{A}$ \cite{jefimenko_electricity_1966,jefimenko_comment_1990,griffiths_timedependent_1991}. We define the corresponding vector potential operator as 
\begin{equation}\label{Jefimenko_Vector_Potential}
\widehat{\mathbf{A}}(\mathbf{r},t)=\frac{\mu_{0}}{4\pi}\int_{\text{all}}\frac{\widehat{\mathbf{j}}\left(\mathbf{r}',t_{r}\right)}{\left|\mathbf{r}-\mathbf{r}'\right|}d\mathbf{r'} ,
\end{equation}
where $t_{r}\equiv t-\left|\mathbf{r}-\mathbf{l}\right|/c$ is the retarded time (with $c$ being the speed of light), properly reflecting the role of the current density as a physical source of the field.  The integration in Eq.~\eqref{Jefimenko_Vector_Potential} is performed over the whole of space and (ignoring the electron spin for simplicity)
\begin{equation}\label{current_operator}
\widehat{\mathbf{j}}(\mathbf{r},t)=-i\left[\left(\nabla_{\mathbf{r}}-\nabla_{\mathbf{r}'}\right)\hat{\Psi}^{\dagger}(\mathbf{r},t)\hat{\Psi}(\mathbf{r}',t)\right]_{\mathbf{r}'\rightarrow\mathbf{r}}
\end{equation}
is the current density operator in the Heisenberg picture directly related to the electronic field operators $\hat{\Psi}(\mathbf{r},t)$. In Eq.~\eqref{current_operator} (and throughout) we consider atomic units where the electron mass is set to unity.

Hence, the vector potential becomes an electronic operator. Correspondingly, the magnetic field is obtained by taking the appropriate quantum statistical-mechanical average of the magnetic field operator $\widehat{\mathbf{B}}(\mathbf{r},t)=\nabla\times\widehat{\mathbf{A}}(\mathbf{r},t)$ to obtain the Jefimenko generalization of the Biot-Savart law to the time-dependent regime,
\begin{align}\label{Bfield_space_integral}
\mathbf{\mathbf{B}\left(\mathbf{r},t\right)} = \frac{\mu_{0}}{4\pi} & \int_{\text{all}} \left[\frac{\mathbf{r}-\mathbf{r}'}{\left|\mathbf{r}-\mathbf{r}'\right|^{3}}\left\langle \widehat{\mathbf{j}}\left(\mathbf{r}',t_{r}\right)\right\rangle \right.\nonumber \\ 
& \left.+ \frac{1}{c}\frac{\mathbf{r}-\mathbf{r}'}{\left|\mathbf{r}-\mathbf{r}'\right|^{2}}\frac{\partial}{\partial t}\left\langle \widehat{\mathbf{j}}\left(\mathbf{r}',t_{r}\right)\right\rangle \right]d\mathbf{r}' .
\end{align}
The required average of the current density operator may then be related to the electronic Green's function
\begin{align}\label{GF_spatial}
& \left\langle \widehat{\mathbf{j}}\left(\mathbf{r},t\right)\right\rangle \nonumber \\
& = -\sum_{\substack{\mu,\nu \\ \in S\cup C\cup D}}\left[\phi_{\mu}(\mathbf{r})\nabla\phi_{\nu}^{*}(\mathbf{r})-\left(\nabla\phi_{\mu}(\mathbf{r})\right)\phi_{\nu}^{*}(\mathbf{r})\right]G_{\mu\nu}^{<}(t,t) ,
\end{align}
where $\left\{ \phi_{\mu}(\mathbf{r})\right\}$ represent an appropriate basis set in which the Green's function is expanded, e.g., atomic orbitals centred on atoms of the molecule and on the leads. Eqs.~(\ref{Bfield_space_integral}) and ~(\ref{GF_spatial}) enable one to obtain the exact magnetic field anywhere in the junction. 

To simplify Eq.~\eqref{Bfield_space_integral}, we replace the spatial integrals in this formula with a line integral along the path comprising the molecular skeleton and the leads. To this end, the integration path is broken up into a set of $N_{s}-1$ segments $P_{n}$ connecting $N_{s}$ sites, and we can define the bond current along each segment $P_{n}$ (from site $n$ to $n+1$) as 
\begin{equation}\label{Bond_Current}
I_{n,n+1}\left(t\right)\equiv4\textrm{Im}\left[T_{n,n+1}\mathbf{\rho}_{n+1,n}\left(t\right)\right] ,
\end{equation}
where the elements on the molecular density matrix are defined as
\begin{equation}\label{density}
\mathbf{\rho}_{n+1,n}\left(t\right)=-i\mathbf{G}_{n+1,n}^{<}\left(t,t\right)
\end{equation}
and the lesser Green's function is given by Eq. (\ref{G_Lesser_Greater}). The bond currents are known to satisfy the relation~\cite{tuovinen_time-dependent_2014}
\begin{equation}\label{Bond_conservation}
\partial_{t}N_{n}\left(t\right)=\underset{m}{\sum}4\textrm{Im}\left[T_{nm}\mathbf{\rho}_{mn}\left(t\right)\right]=\underset{m}{\sum}I_{mn}\left(t\right) ,
\end{equation}
so we can rewrite equation \eqref{Bfield_space_integral} as a summation over the contributions coming from each path $P_{n}$:
\begin{align}\label{BField_Bond}
\mathbf{B}\left(\mathbf{r},t\right)=\frac{\mu_{0}}{4\pi}\underset{n=1}{\overset{N_{s}-1}{\sum}} \int_{P_{n}} & \left[\frac{I_{n,n+1}\left(t_{r}\right)}{\left|\mathbf{r}-\mathbf{l}\right|^{3}} \right.\nonumber \\
& \left.+\frac{1}{c}\frac{\partial}{\partial t}\frac{I_{n,n+1}\left(t_{r}\right)}{\left|\mathbf{r}-\mathbf{l}\right|^{2}}\right]d\mathbf{l}\times\left(\mathbf{r}-\mathbf{l}\right) .
\end{align}

Similarly, we can evaluate the time-dependent electric field~\cite{jefimenko_electricity_1966}
\begin{align}\label{EField_Bond}
& \mathbf{E}(\mathbf{r},t) \nonumber \\
= \frac{1}{4\pi\epsilon_0} & \left\{ \sum_n\left[\frac{\mathbf{r}-\mathbf{r}_n}{|\mathbf{r}-\mathbf{r}_n|^3}\rho_{nn}(t_r) + \frac{\mathbf{r}-\mathbf{r}_n}{|\mathbf{r}-\mathbf{r}_n|^2}\frac{1}{c}\frac{\partial \rho_{nn}(t_r)}{\partial t}\right]\right.\nonumber\\
& \left.-\sum_n \int_{P_n}\frac{1}{c^2}\frac{\partial}{\partial t}\frac{I_{n,n+1}(t_r)}{|\mathbf{r}-\mathbf{l}|}d\mathbf{l}\right\} ,
\end{align}
where $\epsilon_0$ is the vacuum permittivity. In deriving Eq.~\eqref{EField_Bond} we used the fact that in our model the charge distributions are perfectly localized, $\rho(\mathbf{r}',t_r) = \sum_n\rho_{nn}(t_r)\delta(\mathbf{r}',\mathbf{r}_n)$, around the atomic positions $\mathbf{r}_n$. It is worth noting that the Jefimenko equation for the electric field, Eq.~\eqref{EField_Bond}, includes effects from both charge and current densities. Eqs. (\ref{BField_Bond}) and (\ref{EField_Bond}) can be evaluated using Eqs.~\eqref{Bond_Current} and~\eqref{density} and the formula for derivatives of the lesser Green's function:
\begin{align}\label{G_Derivative}
& \frac{\partial\mathbf{G}_{CC}^{<}\left(t,t\right)}{\partial t} =  i\int\frac{d\omega}{2\pi}f\left(\omega-\mu\right) \times \nonumber \\
& \underset{\beta}{\sum} \left[ -i\mathbf{h}_{CC}^{eff}\mathbf{S}_{\beta}\left(t_{r},t_{0};\omega\right)\Gamma_{\beta}\mathbf{S}_{\beta}^{\dagger}\left(t_{r},t_{0};\omega\right) \right. \nonumber\\
& \left. -ie^{-i\omega\left(t_{r}-t_{0}\right)}e^{-i\psi_{\beta}\left(t_{r},t_{0}\right)}\Gamma_{\beta}\mathbf{S}_{\beta}^{\dagger}\left(t_{r},t_{0};\omega\right) + h.c. \right] .
\end{align}

Finally, we may use the derived results for the time-dependent electric and magnetic fields to investigate the energy flux and the radiated power due to the charge and current sources within the molecule. This can be done via the Poynting vector
\begin{equation}\label{SField}
\mathbf{S}(\mathbf{r},t) = \frac{1}{\mu_0}[\mathbf{E}(\mathbf{r},t)\times\mathbf{B}(\mathbf{r},t)] .
\end{equation}
The power radiated into a solid angle $d\Omega$ at distance $R$ in the direction of $\hat{\mathbf{R}}$ is then given by
\begin{equation}\label{eq:power}
dP = d\Omega R^2 \mathbf{S}\cdot\hat{\mathbf{R}} ,
\end{equation}
and the total radiated power is obtained as a surface integral $P=\int dP$, with $d\Omega \equiv \sin \theta d \theta d \phi$ being represented in terms of the polar ($\theta$) and azimuthal ($\phi$) angles, see Fig.~\ref{fig:schematic}.

We emphasize that Eqs.~\eqref{BField_Bond},~\eqref{EField_Bond}, and~\eqref{SField} can be used to study time-resolved electromagnetic fields and fluxes generated by charges and currents in the molecular device described by Eq.~\eqref{Hamiltonian}. As the solution via the TD-LB approach provides a closed analytical expression for the time dependence of the density matrix of the molecular junction and the interface currents, the time-dependent electromagnetic fields can be evaluated without the necessity of numerically propagating individual single-particle orbitals or Green's functions in time.

\section{Numerical Results}\label{sec:results}

\subsection{Transient currents and induced magnetic fields}\label{TD_plots_Sec}

We simulate transport in a benzene molecule described by a single $\pi$-orbital tight-binding model. We set the hopping integral between the nearest neighbor atomic sites $m$ and $n$ as $T_{mn}=-1.0$~a.u. and zero otherwise, cf.~Eq.~\eqref{Hamiltonian}. The molecule is contacted from two sites to two metallic electrodes ($\alpha=\{S,D\}$) with a sudden voltage drop $V_S(t)=-V_D(t)\equiv V\theta(t-t_0)$ and we set the switch-on time to $t_0=0$, cf.~Eq.~\eqref{TD_energies}. The overall bias window is therefore $V_S-V_D=2V$. We consider two cases of weak ($V=0.1$~a.u.) and strong ($V=1.0$~a.u.) bias. Only the coupling matrix elements between the nearest sites of the electrodes and the molecule are set to nonzero values. $T_{1kS}$ labels the coupling to the source lead (see numbering of sites in Fig.~\ref{fig:schematic}) and $T_{mkD}$ labels the three possible configurations of the drain lead, where $m=4$ in the para configuration, $m=5$ in the meta configuration, and $m=6$ in the ortho configuration.
The energy scale in the electrodes is described by a hopping integral $T_\alpha$, defining the tunneling rate $\Gamma_\alpha=2T_{mk\alpha}^2/|T_\alpha|$ such that the level width matrix in Eq.~\eqref{Gamma_matrix} has the structure:
\begin{equation}\label{Gamma_matrix_benzene}
\Gamma_{\alpha,ij}=\Gamma_{S}\delta_{\alpha S}\delta_{ij}\delta_{i1}+\Gamma_{D}\delta_{\alpha D}\delta_{ij}\delta_{im} .
\end{equation}
We expect the WBLA to be a good approximation for the embedding self-energy when $|T_\alpha| \gg |T_{mk\alpha}|$~\cite{ridley_lead_2019-1}. We further set the inverse temperature to $\beta=100$~a.u.$^{-1}$ and assume symmetry in the lead-molecule couplings, $\Gamma_{S}=\Gamma=\Gamma_{D}$. This implies equal couplings of both leads to the molecular sites, $T_{1kS}=T_{mkD}$.

\begin{figure}[t]
    \centering
    \includegraphics[width=0.5\textwidth]{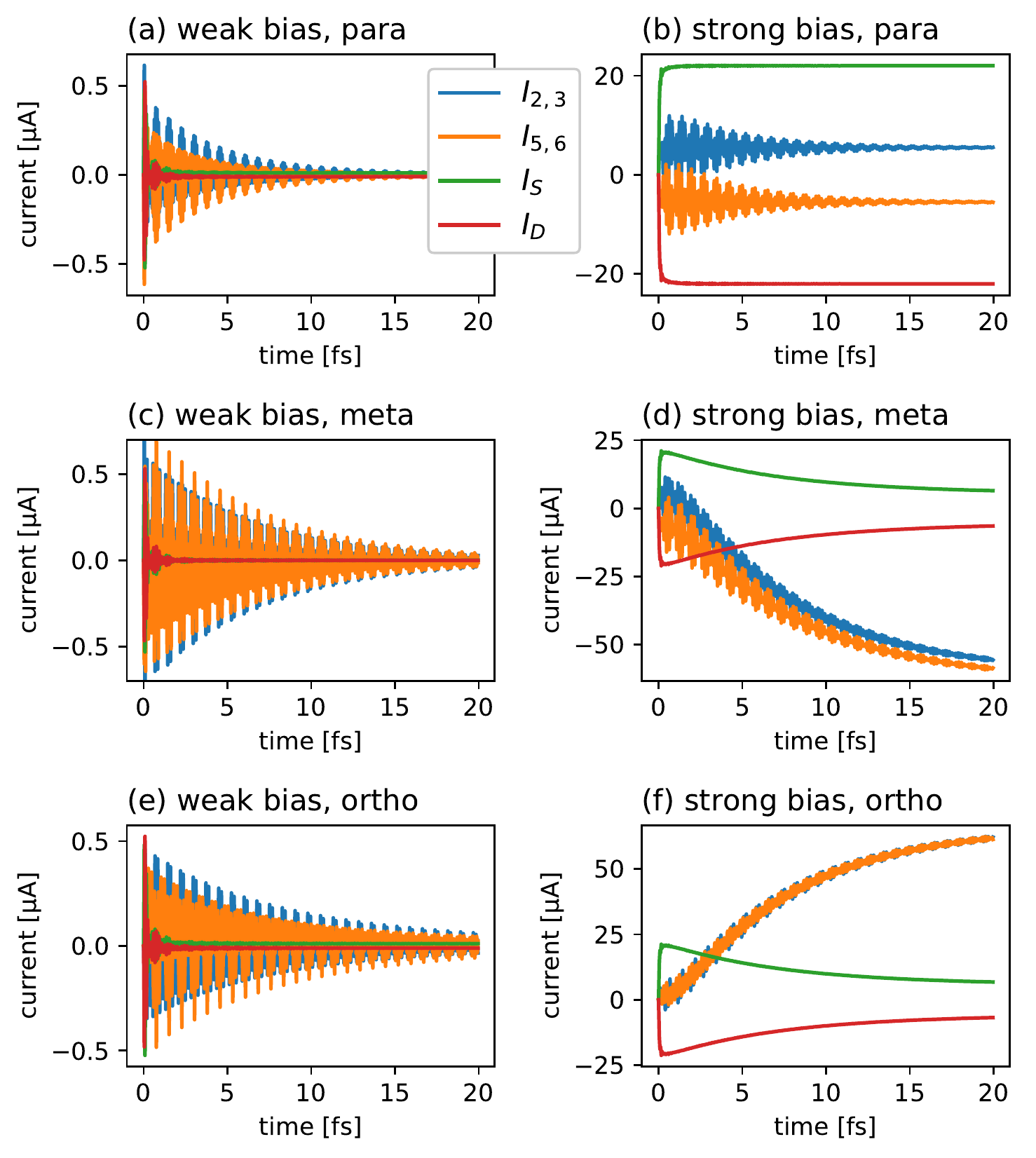}
    \caption{Time-dependent currents through the benzene molecule due to weak bias when $V=0.1$~a.u. (left panels) and strong bias when $V=1.0$~a.u. (right panels), in the para (top panels), meta (middle panels), and ortho (bottom panels) configuration.}
    \label{fig:tdcurrent}
\end{figure}

We address the transient response of the molecular junction by evaluating the time-dependent bond currents from the off-diagonal elements of the density matrix, $\rho(t)=-i G^<(t,t)$, using Eqs.~\eqref{Bond_Current},~\eqref{density} and~\eqref{G_Lesser_Greater}. We also evaluate the time-dependent current at the electrode interface, $I_\alpha(t)$, using Eq.~\eqref{Current_alpha}. Since we also wish to address electromagnetic fields generated by the current sources within the molecular junction, we convert from the atomic units to SI units for current by $1$~a.u.~$\approx 6.624\cdot 10^{-3}$~A, and time $1$~a.u.~$\approx 2.419\cdot 10^{-17}$~s. Figure~\ref{fig:tdcurrent} shows the time-dependent bond currents and the interface currents in case of weak and strong bias for the para, meta, and ortho coupling configurations. In this calculation we have set $T_\alpha=-8.0$~a.u. and $T_{mk\alpha}=-0.2$~a.u. resulting in $\Gamma=0.01$~a.u., i.e., we are justified in using the WBLA. 

We see in Fig.~\ref{fig:tdcurrent} that the currents generated are in the micro-ampere regime, and they saturate towards stationary values in a few tens of femtoseconds. While the interface current behaves rather regularly, the individual bond currents within the molecule exhibit a considerable oscillatory character. These oscillations characterise the timescales of circular currents in the ring-shaped molecule. Interestingly, in the meta and ortho coupling configuration, the electrons within the molecule seem trapped in a highly nonequilibrium state with persistent oscillations even exceeding the total current to the electrodes, see Figs.~\ref{fig:tdcurrent}(d) and~\ref{fig:tdcurrent}(f). In addition, the overall relaxation of the bond currents in the meta and ortho cases is slower than for the para configuration. We attribute this to the longer traversal pathways, and therefore longer timescales for traversal of electrons across the junction in the meta and ortho cases \cite{ridley_electron_2019}. In addition, we observe that the interface current is comparable in strength to the individual bond currents, so both types of current should be taken into consideration when calculating the electromagnetic fields originating from these current sources. It is worth noticing that the interface current is positive from the source electrode to the molecular region, $I_S>0$, while it is negative from the drain electrode to the molecular region, $I_D<0$, i.e., the current is flowing from the source to the drain.

The direction of the current can also be appreciated from the bond currents $I_{2,3}$ and $I_{5,6}$. Since the bond currents are of opposite sign in the para configuration, see Fig.~\ref{fig:tdcurrent}~panels~(a)-(b), there are two current pathways through the molecule. Due to the symmetry of the transport setup, cf.~Fig.~\ref{fig:schematic}, the individual bond currents for the para configuration satisfy $I_{12}=I_{34}\neq I_{23} > 0$ and $I_{45}=I_{61}\neq I_{56} < 0$. On average, the bond currents cancel each other out, and there is no circular current for the para configuration. As seen in Fig.~\ref{fig:tdcurrent}~panels~(c)-(f) this situation changes for the meta and ortho configurations. For the meta configuration, we have $I_{12}=I_{45}\neq I_{23}=I_{34} \neq I_{56} = I_{61} < 0$ due to symmetry, i.e. on average there is a negative bond current which means a circular current in the counter-clockwise direction. For the ortho configuration, in turn, $I_{12}=I_{56} \neq I_{23}=I_{45} \neq I_{34} \neq I_{61} > 0$, i.e. the circular current is in the clockwise direction. We will investigate the circular currents in more detail in Sec.~\ref{sec:interference}.

For the calculation of the time-dependent fields in Eqs.~(\ref{BField_Bond}) and~(\ref{EField_Bond}), we need the spatial coordinates of the atomic sites, between which the bond currents are calculated, for the parametrization of the line integral. We define these coordinates as $\mathbf{r}_j \equiv (x_j,y_j,z_j)$ for $j\in\{1,\ldots,6\}$ and specify the hexagonal structure with lattice constant given by $a=1.4$~\AA. The point at which the magnetic field is calculated, $\mathbf{r} \equiv (x,y,z)$, is constant in the line integral. We parametrize the path between the atomic sites $j$ and $k$ as $\mathbf{l}_{jk} = (1-\tau)\mathbf{r}_j+\tau\mathbf{r}_k$ where $\tau\in[0,1]$. Applying the chain rule we then get 
\begin{equation}
\int_{P_{jk}} \dots d\mathbf{l}_{jk} = \int_0^1 \dots \frac{d\mathbf{l}_{jk}}{d\tau}d\tau = \int_0^1 \dots (\mathbf{r}_k - \mathbf{r}_j)d\tau
\end{equation}
for each segment of the path. As we are dealing with a planar molecule, we set the molecular coordinates in the $z$ direction to zero (see Fig.~\ref{fig:schematic}). This simplifies the calculation of the cross product in Eq. (\ref{BField_Bond}). However, the approach is completely valid for non-planar molecules as long as full three-dimensional lattice coordinates are employed.

The individual bond currents and their time derivatives in Eq. (\ref{BField_Bond}) are to be evaluated at the retarded time, $t_r=t-|\mathbf{r}-\mathbf{l}|/c$. In atomic scale junctions, such as those considered in this work, we have checked the difference between the retarded time and the measurement time to be maximally of the order of attoseconds, falling well below the relevant time scales observed in the time-dependent currents in Fig.~\ref{fig:tdcurrent}. 

\begin{figure}[ht!]
    \centering
    \includegraphics[width=0.375\textwidth]{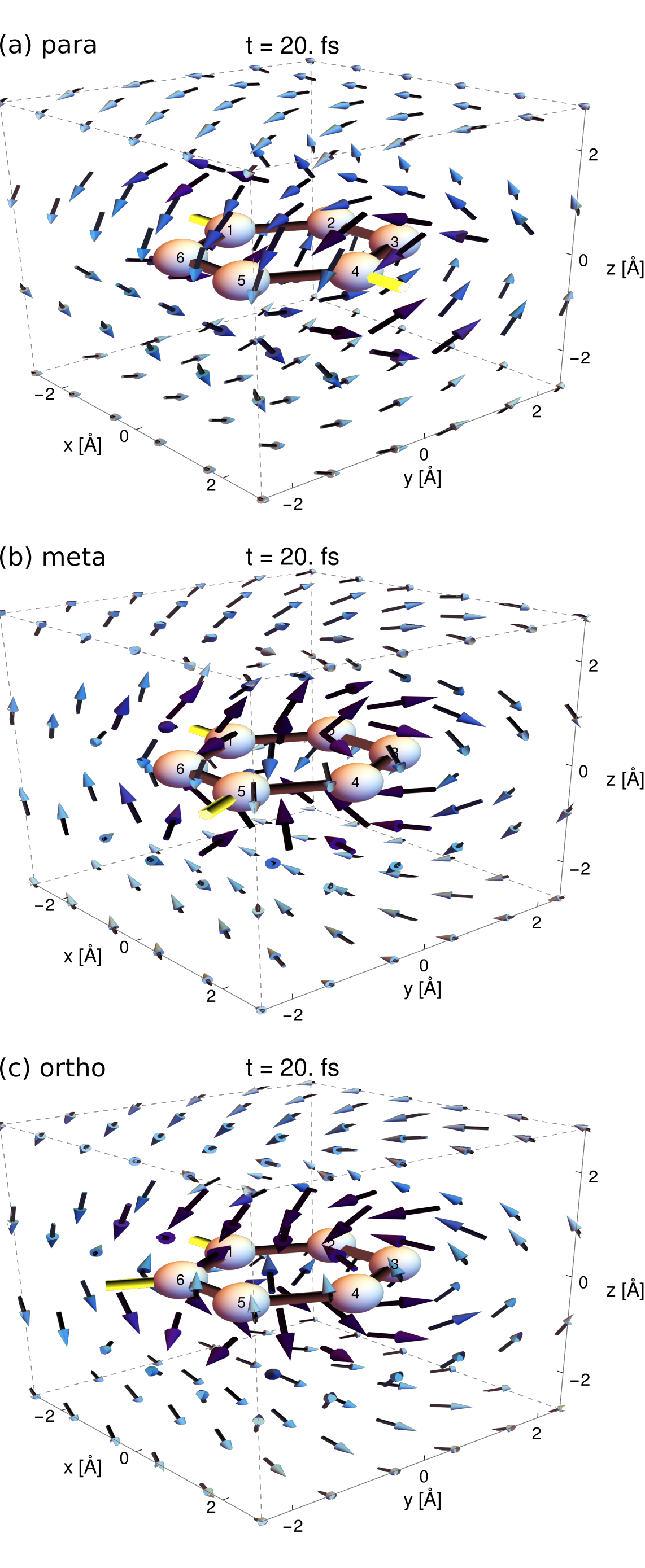}
    \caption{Induced magnetic fields due to strong bias at $t=20$~fs for (a) para, (b) meta, and (c) ortho configuration. The field strength is indicated by the size and color (darkness) of the arrows. The maximum absolute values for the field strengths are (a) $3.5$~mT, (b) $5.6$~mT, and (c) $6.8$~mT.}
    \label{fig:bfield1}
\end{figure}

We now take the results of the calculation of the time-dependent bond currents and the interface currents reported in Fig.~\ref{fig:tdcurrent}, and show the induced magnetic field response to a strong bias in Fig.~\ref{fig:bfield1} for the para, meta and ortho coupling configurations. We take the interface currents into account by adding dangling bonds in the corresponding direction of the hexagonal lattice (see the yellow bonds in Fig.~\ref{fig:bfield1}), where Eq.~\eqref{Current_alpha} is used to compute the current. This is because evaluating Eq.~\eqref{BField_Bond} requires a vector $\mathbf{l}$ at which the current is flowing, so we must calculate the line integral along these dangling bond paths. The direction of this bond is specified by the hexagonal lattice unit vectors, and it changes depending on the coupling configuration. We first analyze the structure of the magnetic vector field at one instant of time at $t=20$~fs.

At $t=20$~fs the individual bond currents have mostly saturated to their stationary value. This means that the induced magnetic field at this instant of time is also stationary like its current sources. In the para configuration [Fig.~\ref{fig:bfield1}(a)] the current flows in one direction globally, and the molecule acts more as a current carrying wire. The induced magnetic field respects this symmetry and shows an overall circular form around the molecule in the transport direction. The transport setup in the meta and ortho configurations is asymmetrical. Therefore, also the induced magnetic field is asymmetrical with respect to the molecular geometry, as shown in Figs.~\ref{fig:bfield1}(b) and~\ref{fig:bfield1}(c). Interestingly, as we already saw in Fig.~\ref{fig:tdcurrent}, for the ortho case the first site of the molecule is coupled to the source lead and the sixth site to the drain lead, and electronic motion is observed in the clockwise direction. This fact can be appreciated by the apparent right-hand rule for the direction of the induced magnetic field in Fig.~\ref{fig:bfield1}(c). In the meta configuration, the current direction is opposite to the ortho case, and this is also observed in the induced magnetic field structure in Fig.~\ref{fig:bfield1}(b). Overall, the strength of the induced magnetic field is of the order of milli-Tesla in this case of strong bias and weak coupling.

\begin{figure}[t]
    \centering
    \includegraphics[width=0.5\textwidth]{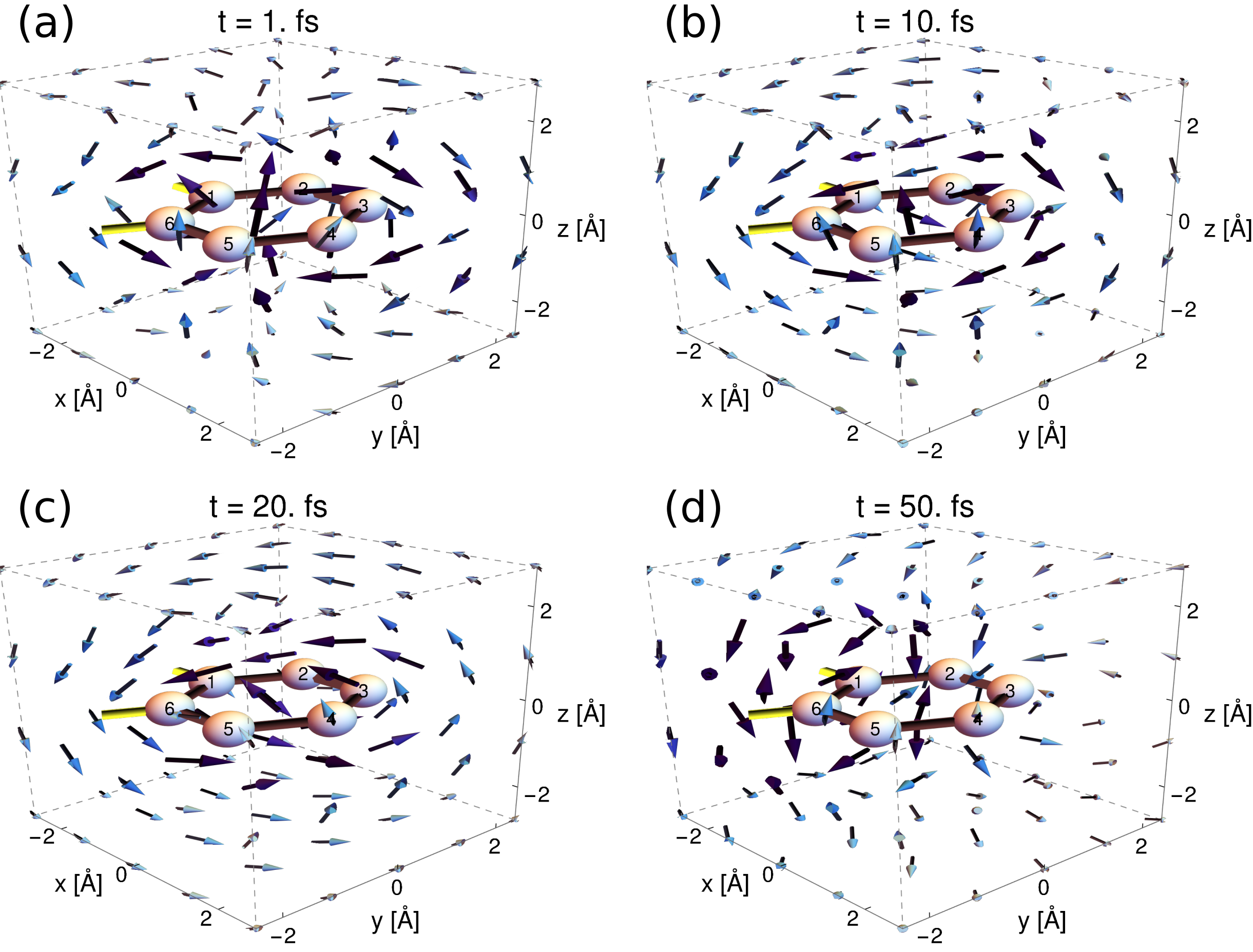}
    \caption{Snapshots of induced magnetic fields due to weak bias at different instants: (a) $t=1$~fs, (b) $t=10$~fs, (c) $t=20$~fs, and (d) $t=50$~fs. The molecule is coupled to the electrodes in the ortho configuration. The maximum absolute value for the field strength is (in chronological order)
    $53 \ \mu$T, $14 \ \mu$T, $10 \ \mu$T, and $3.2 \ \mu$T.
    }
    \label{fig:bfield2}
\end{figure}

We display the temporal evolution of the magnetic field in Fig.~\ref{fig:bfield2}. This corresponds to the weak bias case in the ortho configuration, cf.~Fig.~\ref{fig:tdcurrent}(e). At the initial transient ($t \lesssim 10$~fs) the magnetic field is strongly localized around the sites of the molecule lying opposite to the electrodes. This is due to strong focusing of the individual bond currents causing the vortical structure for the induced magnetic field. As the currents start to relax towards their stationary values ($t \gtrsim 20$~fs) the induced magnetic field first becomes more delocalized around the whole molecule, and ultimately ($t \gtrsim 50$~fs) focuses again around the strongest currents at the molecule-lead interface. In this case of weak bias and coupling, the induced magnetic field strengths are in the micro-Tesla regime.

\subsection{Steady state interference effects}\label{sec:interference}

\begin{figure}[t]
    \centering
    \includegraphics[width=0.475\textwidth]{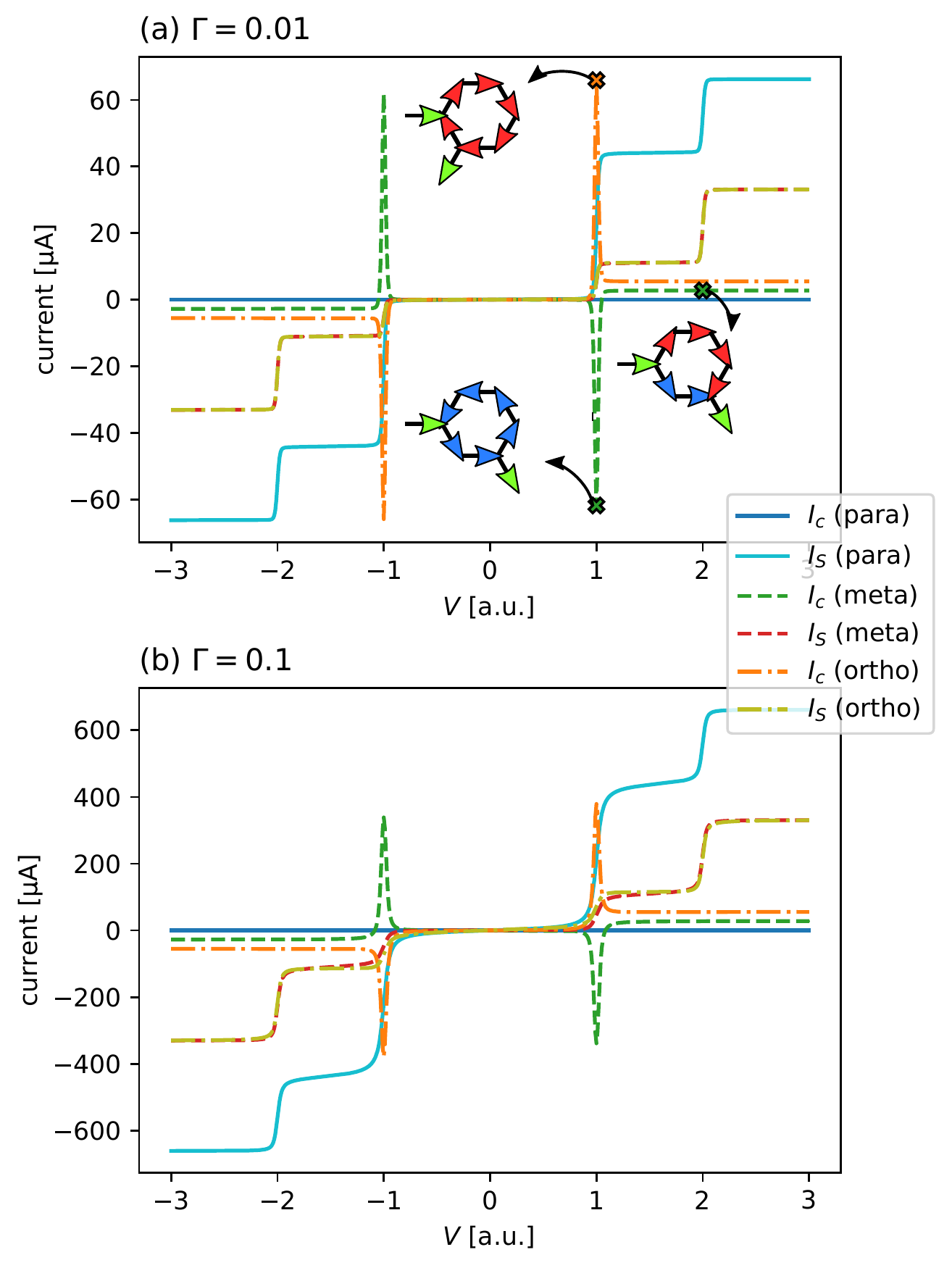}
    \caption{The steady state current response to the bias for para, meta and ortho configurations at (a) $\Gamma=0.01$ and (b) $\Gamma=0.1$. We show the circular current $I_c$, defined as the average of the individual bond currents, and the interface current $I_S$, evaluated for different voltages at the steady state $t\to\infty$. Insets in panel (a) show the individual bond-current orientation at specific voltages. The coloring of the arrows in the insets is green for the interface currents, and red (blue) for positive (negative) circular current.}
    \label{fig:I_SS}
\end{figure}

In this section we will focus on the long-time limit of the transient observables, i.e., we look at the steady state currents and induced magnetic fields at $t \to \infty$. In addition to the weak-coupling case in the previous subsection ($\Gamma=0.01$~a.u.) we now also look at the intermediate coupling regime (with respect to the intra-molecular hopping) with $T_\alpha=-5.0$~a.u. and $T_{mk\alpha}=-0.5$~a.u. resulting in the tunneling rate $\Gamma=0.1$~a.u. In Fig.~\ref{fig:I_SS} we show the steady-state current-voltage characteristics for the setting considered in Fig.~\ref{fig:tdcurrent} for the interface current $I_S$ and for the net circular current, which we define as the average bond current $I_c = 1/N \sum_i I_{i,i+1}$, where $N=6$ represents the number of sites in the molecule. As a general observation, a notable current starts flowing through the benzene molecule only when the highest-occupied and lowest-unoccupied molecular orbitals (HOMO and LUMO) at energies $\pm |T_{mn}| = \pm 1.0$~a.u. are included in the bias voltage window. We note that the uncontacted benzene ring has degenerate energy eigenvalues at $\pm\left|T_{mn}\right|$ in addition to two non-degenerate eigenstates at $\pm2\left|T_{mn}\right|$ \cite{hansen_interfering_2009}. The current response is more smeared out when $\Gamma$ is increased due to the correspondingly broader spectrum of the unstable eigenmodes of the molecular structure, cf.~Eq.~\eqref{Effective_Hamiltonian}.

The interface current in Fig.~\ref{fig:I_SS}(a) has steps at the eigenvalues of the benzene ring, due to the approximately arctangent shape of the $I-V$ characteristic at low temperatures \cite{ridley_current_2015}. The steps gain a slope in Fig.~\ref{fig:I_SS}(b) at the stronger coupling $\Gamma=0.1$, due to corresponding broadening of the transmission probability at molecular eigenvalues with increasing $\Gamma$. As expected, the circular current in the para configuration is zero independently of the coupling strength and for all voltages due to the complete symmetry of the para transport setup. By contrast, at $|V|=1.0$~a.u. there is a strong resonance in the circular currents for the meta and ortho configurations, see the insets in Fig.~\ref{fig:I_SS}(a). At this voltage, the circular currents even exceed the interface current, cf.~Fig.~\ref{fig:tdcurrent}. For voltages higher than the resonant voltage the circular currents decrease in strength to a saturated value which is larger for the ortho configuration. At this regime, there are two current pathways through the molecule also for the meta and ortho cases. For $V$ values equal to the subsequent energy eigenvalues at $\pm 2.0$~a.u., the interface current increases to a saturated value as there are no more transport channels available at higher voltages. Interestingly, the circular current is unaffected by the additional transport channel at $|V|=2.0$~a.u. The individual bond currents do change at $|V|=2.0$~a.u. but the change occurs in opposite directions so that the circular current remains unchanged, see the insets in Fig.~\ref{fig:I_SS}(a). 

To help interpret our results, we can analytically evaluate the source current (Eq.~(\ref{Current_alpha})) and bond current (Eq.~(\ref{Bond_Current})) in the steady state limits:
\begin{widetext}
\begin{align}
I_{S} & = 2\Gamma^{2}\int\frac{d\omega}{2\pi}f\left(\omega-\mu\right)\left[\left|G_{1m}^{r}\left(\omega+V\right)\right|^{2}-\left|G_{1m}^{r}\left(\omega-V\right)\right|^{2}\right] , \label{I_S} \\
I_{n,n+1} & = 4\Gamma\int\frac{d\omega}{2\pi}f\left(\omega-\mu\right)\textrm{Im}\left[T_{n,n+1}\left(G_{n+1,1}^{r}\left(\omega+V\right)G_{1,n}^{a}\left(\omega+V\right) +  G_{n+1,m}^{r}\left(\omega-V\right)G_{m,n}^{a}\left(\omega-V\right)\right)\right] . \label{I_nnplus1}
\end{align}
\end{widetext}
Here we have used Eq.~\eqref{Gamma_matrix_benzene} to express the frequency integrands as a scalar product of Green's function components, where the indices 1 and $m$ denote those sites on the benzene ring which are coupled to the $S$ and $D$ electrodes, respectively. This simplification helps us to explain the antisymmetry in $V$ -- making the replacement $V \rightarrow -V$ in the integrands of Eq.~\eqref{I_S} or Eq.~\eqref{I_nnplus1} is equivalent to switching the indices 1 and $m$. Due to the particle-hole symmetry for the model of the benzene molecule, this is equivalent to reversing the direction of the currents.

In Refs. \cite{solomon_understanding_2008,hansen_interfering_2009,markussen_relation_2010} the impact of quantum interference on the transport is analysed in terms of the the zeros of $G_{1m}^{r}$, which can be obtained from a cofactor matrix method and which give zero transmission probability in Eq. (\ref{I_S}). However, the bond current in Eq. \eqref{I_nnplus1} depends on different components in the Green's function, namely $G_{n+1,1}^{r}$, $G_{1,n}^{a}$, $G_{n+1,m}^{r}$ and $G_{m,n}^{a}$. The two terms in the integrand of Eq. \eqref{I_nnplus1} represent a superposition of bond current densities at the source and drain. The interference between these two terms explains the resonances shown in Fig. \ref{fig:I_SS}. In the perfectly symmetrical case of the para configuration, there is a total cancellation of terms in the integrand which kills the resonant peak. We include plots of the integrands of Eqs.~\eqref{I_S} and~\eqref{I_nnplus1} around the resonant voltage in the supplemental material~\cite{sm} for additional verification of this effect.

\begin{figure}[t]
    \centering
    \includegraphics[width=0.475\textwidth]{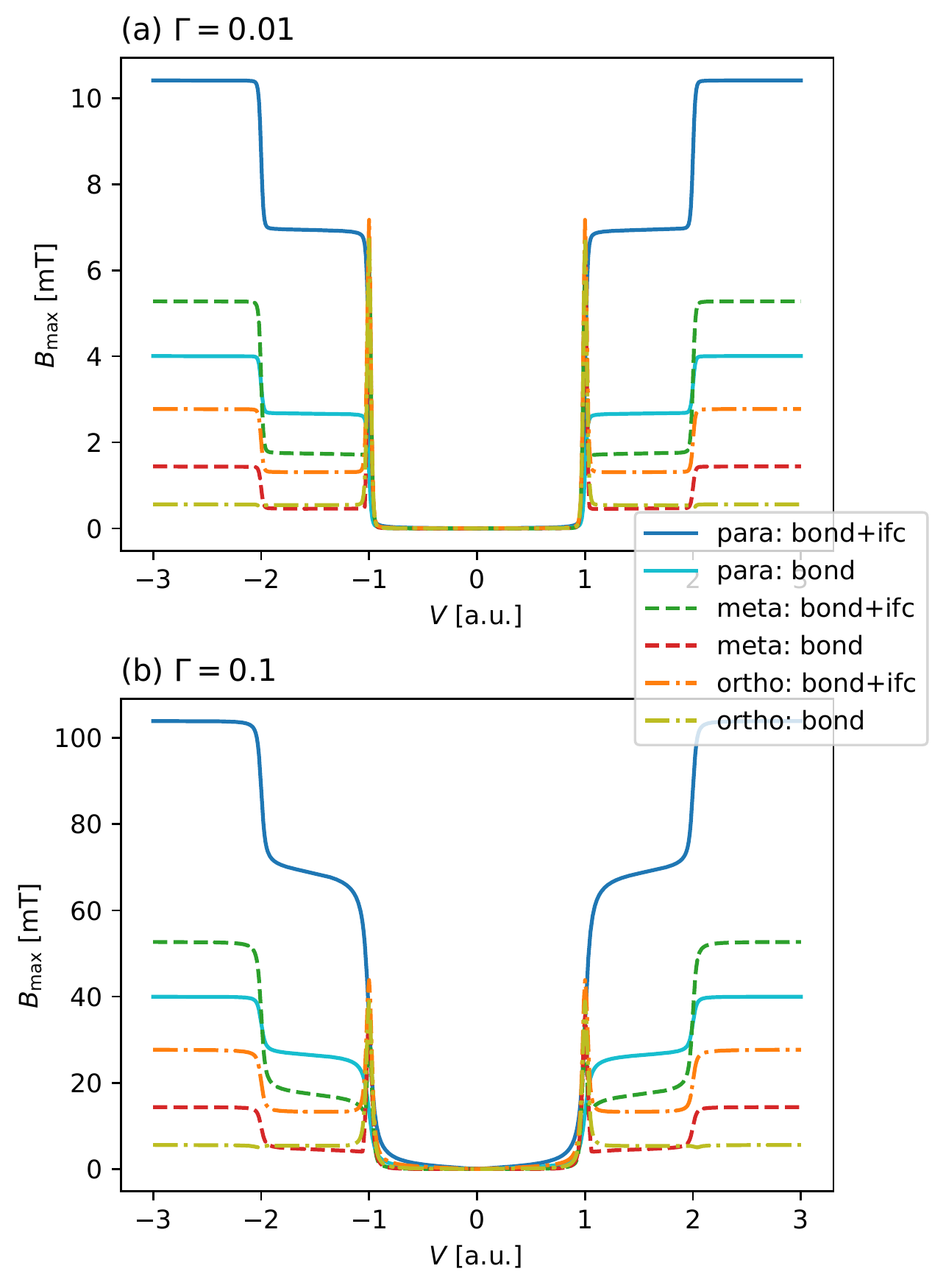}
    \caption{The maximum magnetic field response to the bias for para, meta and ortho configurations at (a) $\Gamma=0.01$ and (b) $\Gamma=0.1$. We show the magnetic field resulting from the bond currents alone (bond), and from the bond and interface currents combined (bond+ifc), cf.~Fig.~\ref{fig:I_SS}.}
    \label{fig:B_max}
\end{figure}

The structure of the current-voltage characteristics is naturally reflected in the induced magnetic field. In Fig.~\ref{fig:B_max} we show the absolute value of the maximum induced magnetic field around the molecular junction for the same setting as in Fig.~\ref{fig:I_SS}. Importantly, for the meta and ortho configurations, the resonance at $V=\pm 1.0$~a.u. is also clearly visible in the induced magnetic field. Similarly to the current response, the resonance is not as sharp when $\Gamma$ is increased. In Fig.~\ref{fig:I_SS} we observed that the circular current is unaffected by the additional transport channel at $|V|=2.0$~a.u., but the individual bond currents vary at this point. This can be seen in the induced magnetic field which is affected by the individual bond currents even if the circular current (their average) remains the same. 

\subsection{The quantum emitter}\label{sec:radiation}

\begin{figure*}[t]
    \centering
    \includegraphics[width=0.85\textwidth]{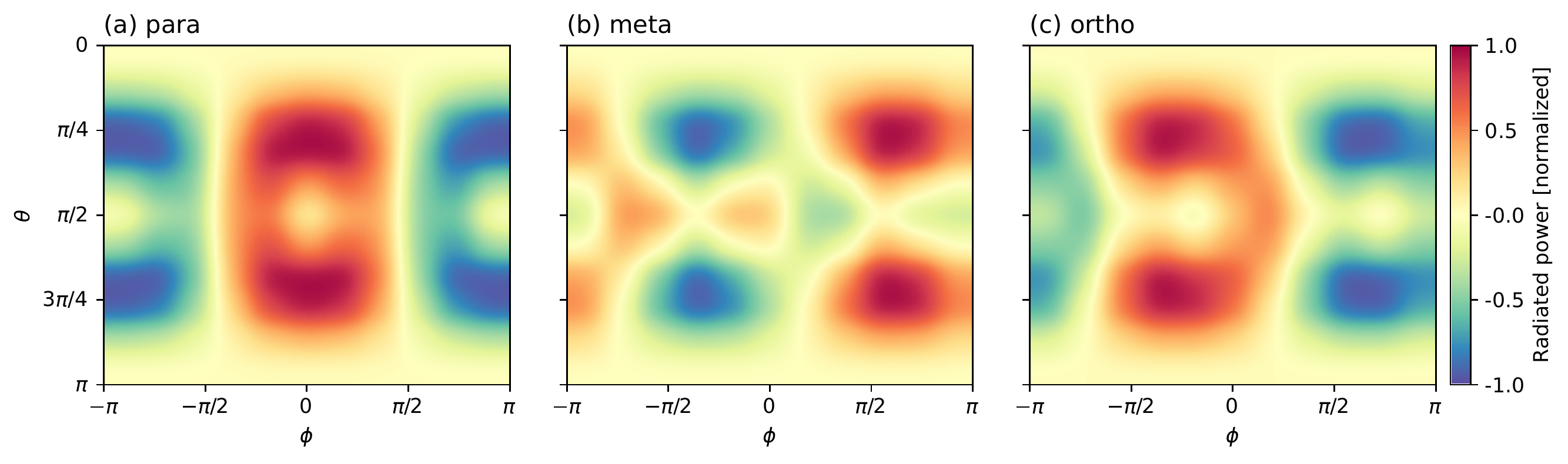}
    \caption{Stationary radiated power into a solid angle $d\Omega$ at distance $R=10d$, where $d=2a=2.8$~\AA~is the molecular diameter, for (a) para, (b) meta, and (c) ortho coupling configurations. The horizontal and vertical axes represent the azimuthal ($\phi$) and polar ($\theta$) angles, respectively. The bias voltage is set by $V=1.0$~a.u.}
    \label{fig:radiationV1}
\end{figure*}

\begin{figure*}[t]
    \centering
    \includegraphics[width=0.85\textwidth]{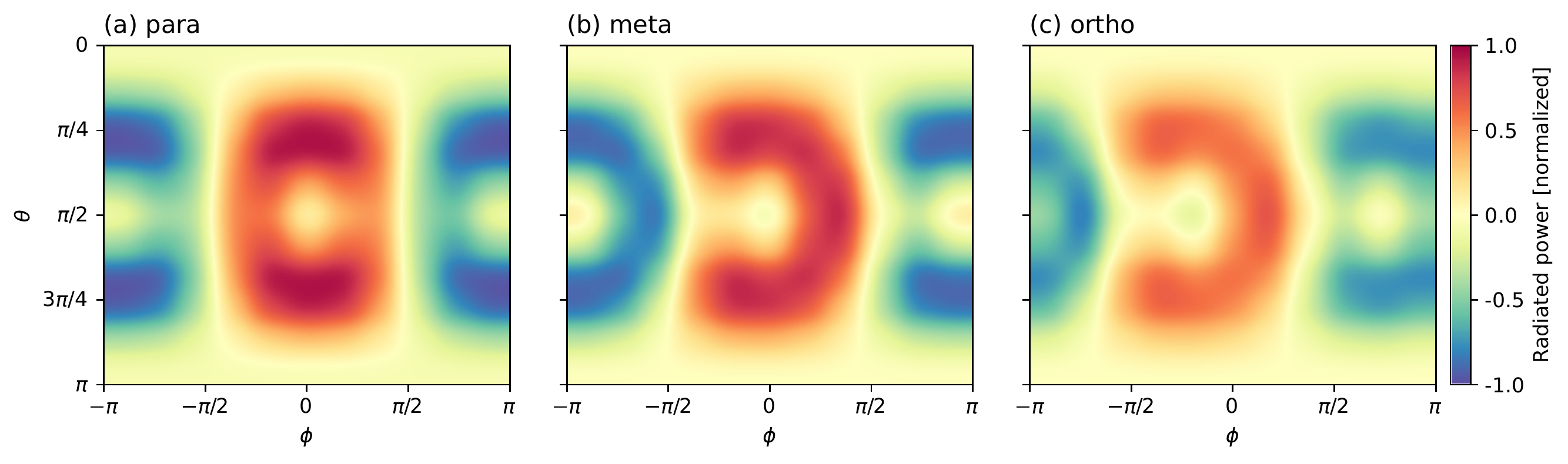}
    \caption{Same as Fig.~\ref{fig:radiationV1} but with the bias voltage set by $V=2.0$~a.u.}
    \label{fig:radiationV2}
\end{figure*}

Finally, we consider the three-dimensional radiation pattern around the molecular device, represented in terms of the polar ($\theta$) and azimuthal ($\phi$) angles, with the origin of coordinates located at the center of the benzene ring, see Fig.~\ref{fig:schematic}. We evaluate the radially emitted power using Eq.~\eqref{eq:power} at distance $R=10d$ where $d=2a=2.8$~\AA~is the molecular diameter. The full angular profile of the radiation flux for the three molecular configurations is shown in Figs.~\ref{fig:radiationV1} and~\ref{fig:radiationV2}. All of the junction parameters in these plots are the same as in Fig.~\ref{fig:I_SS} (a), and all fields are calculated in the steady state limit. We include the full three-dimensional vector field plots of $\textbf{E}$, $\textbf{B}$ and $\textbf{S}$ for the different configurations in 
the supplemental material~\cite{sm}. It is immediately apparent from inspection of these plots that the electric field is not strongly dependent on the molecular configuration, whereas the magnetic field is a much more sensitive probe of the local currents. Thus the differences in radiation flux may be attributed to differences in the magnetic field for the different molecular geometries. 

We now consider the angular profile of the radiation flux at the resonant point $V=1.0$~a.u., shown in Fig.~\ref{fig:radiationV1}. Firstly, all plots show zero flux in the $\theta=0,\pi$ direction perpendicular to the plane of the molecule. The para configuration exhibits maxima at $\pm 45^\circ$ to the plane, with positive and negative maxima in the energy flux along the $\phi=0,\pm\pi$ directions, respectively. Thus there is a direct correspondence between the azimuthal direction of current and energy flux, even though the relative dependence on the polar angle is offset. The radiated power is therefore extremely sensitive to the reversal in the sign of the circular current. This explains the inversion in the direction of radiated power seen in the meta configuration [Fig.~\ref{fig:radiationV1}(b)] relative to the para and ortho cases, and the inversion of the sign on the flux about the point $\theta=\pi/2$, $\phi=\pi/6$ in this plot. This corresponds to the angle at which the drain lead is bonded to the benzene molecule, where there is a switch in direction of circular current flow. 

In Fig.~\ref{fig:radiationV2}, the radiated power is shown for $V=2.0$~a.u. As shown in the inset to Fig.~\ref{fig:I_SS} (a), the circular currents in the meta configuration are now running from source to drain, which means the direction of power flux is reversed relative to the $V=1.0$~a.u. case. We note that for both voltages, additional hot (or cold) spots are seen at the azimuthal angles $\phi=-2\pi/3,\pi/6$ in the meta configuration and $\phi=-5\pi/6,\pi/3$ in the ortho configuration. These are the bonding angles to the drain leads. This observation provides a clear signature of the molecular geometry irrespective of the voltage.

We note that Figs.~\ref{fig:radiationV1} and~\ref{fig:radiationV2} show regions with negative radiated power. These correspond to the situation when the Poynting vector is directed towards the molecule, see also the supplemental material~\cite{sm}. We expect that in the case of far-field radiation, the radiated power patterns would reshape in such a way that there would be no negative areas. However, in our nanojunction setting, the generated fields are fairly weak. As their amplitudes also decay significantly over distance, cf.~Eqs.~\eqref{BField_Bond} and~\eqref{EField_Bond}, we do not find a large enough energy flux at infinity to address this issue numerically. Nevertheless, the integrated power through any spherical surface around the molecule is a constant due to energy conservation.

\section{Conclusions}

We have studied time-resolved electromagnetic fields due to transient current sources in molecular junctions. We employed Jefimenko's retarded solutions to the Maxwell equations together with the TD-LB approach to obtain a closed analytical expressions for $\mathbf{B}(\mathbf{r},t)$ and $\mathbf{E}(\mathbf{r},t)$. Owing to the TD-LB approach, which is well-supported by the underlying nonequilibrium Green's function theory, the methodology we have presented offers a fast and accurate way of addressing macroscopically emergent effects caused by microscopic quantum transport phenomena out of equilibrium.

We applied the formalism to a benzene-molecule junction coupled to electrodes in different coupling geometries. We found symmetry-driven interference effects for the transient current behavior which, in turn, translated to a detailed temporal relaxation of the induced magnetic field in the vicinity of the molecule. To address the quantum interference effects, we also investigated the stationary current-voltage characteristics where we found novel resonances in the ortho and meta coupling configurations. The resonant behavior, which can be identified with interfering electron pathways through the molecule, was also discovered in the induced magnetic field. Finally, we calculated the angular dependence of power radiated from the benzene molecule for the different geometries. We found strong non-local signatures of the type of molecular coupling in the resonant regime, where quantum interference has a qualitative effect on the radial flow of electromagnetic energy density. 

We have concentrated on a noninteracting picture although many-body correlations could, in principle, influence the transport mechanisms leading to the local radiation profile of the molecule~\cite{ridley_numerically_2019}. For example, it would be informative to study the image-charge effect, which can affect the HOMO-LUMO gap and corresponding charge oscillations~\cite{Myohanen2012}. We would expect this to alter the local electromagnetic fields in a nontrivial way. These effects could be addressed, e.g., using the generalized Kadanoff-Baym ansatz for open quantum systems~\cite{Tuovinen2020b}.

In future work, we will develop radiation profiles for different molecular structures using the method detailed here. Work done on nanoantennas to date has focused on the impedance and frequency response to external time-dependent driving fields \cite{greffet_impedance_2010}, where additional quantum effects such as photon-assisted tunneling become important. We have already used the TD-LB approach to investigate the effects of such driving on the current \cite{ridley_time-dependent_2017} and noise \cite{ridley_partition-free_2017}, but the non-local features of the surrounding electromagnetic fields can also be mapped for arbitrary time-dependent external biases. 

\begin{acknowledgments}
This work has been supported in part by the Israel Science Foundation Grant No. 2064/19 and the National Science Foundation--US-Israel Binational Science Foundation Grant No. 735/18 (M.R.), and by the Academy of Finland Project No. 321540 (R.T.). R.v.L. likes to thank the Academy of Finland for support under Grant No. 317139.
\end{acknowledgments}


%

\clearpage
\widetext

\begin{center}
\textbf{\large Supplemental Material: Quantum interference and the time-dependent radiation of nanojunctions}
\end{center}
\setcounter{equation}{0}
\setcounter{figure}{0}
\setcounter{table}{0}
\setcounter{page}{1}
\makeatletter
\renewcommand{\theequation}{S\arabic{equation}}
\renewcommand{\thefigure}{S\arabic{figure}}
\renewcommand{\thepage}{S\arabic{page}}
\renewcommand{\bibnumfmt}[1]{[S#1]}
\renewcommand{\citenumfont}[1]{S#1}

Here we present additional figures supporting the findings reported in the main text.

\begin{figure}[h!]
    \centering
    \includegraphics[width=0.85\textwidth]{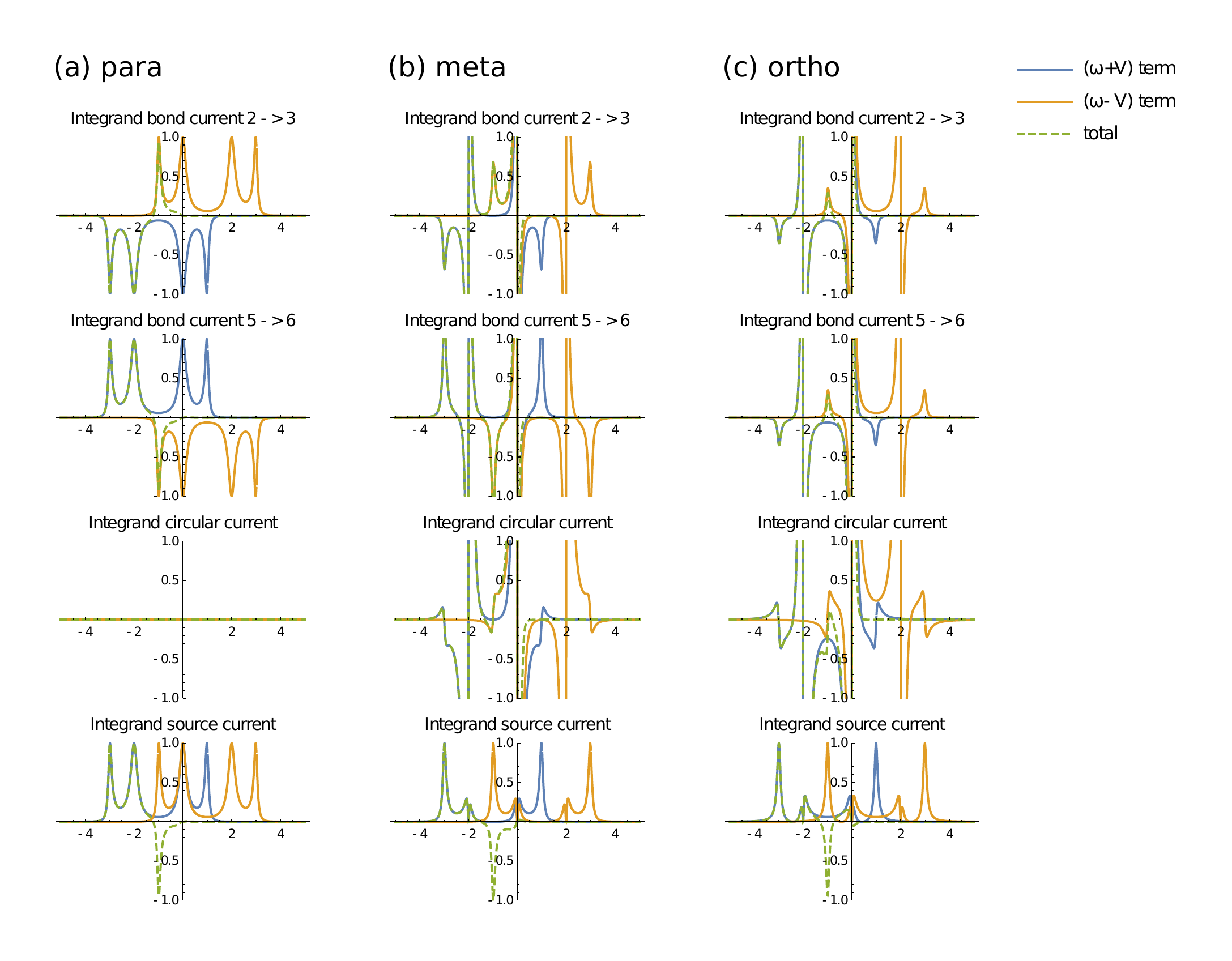}
    \caption{Frequency dependency of the integrands in Eqs.~\eqref{I_S} and~\eqref{I_nnplus1} at the resonant voltage $V=1.0$~a.u. for (a) para, (b) meta, and (c) ortho coupling configurations. The two curves, `$\omega+V$' and `$\omega-V$', represent the two terms in the square brackets of the integrands. The `total' curve is the sum of the two terms including the Fermi function. For the sake of better visualization, we have set $\Gamma=0.5$ and $\beta=10$ so that the peaks are wider and the Fermi function behaviour can be seen better around zero. This visualization choice does not affect the conclusions drawn in the main text. For meta and ortho the terms interfere constructively around $\omega=0$ enhancing the bond and circular current. We note, however, that the Fermi function can kill the interference for positive $\omega$ particularly at higher temperatures. For the para case, the terms interfere destructively leading to the total cancellation in the calculation of the circular current.}
    \label{fig:integrands}
\end{figure}

\begin{figure}[h!]
    \centering
    \includegraphics[width=0.85\textwidth]{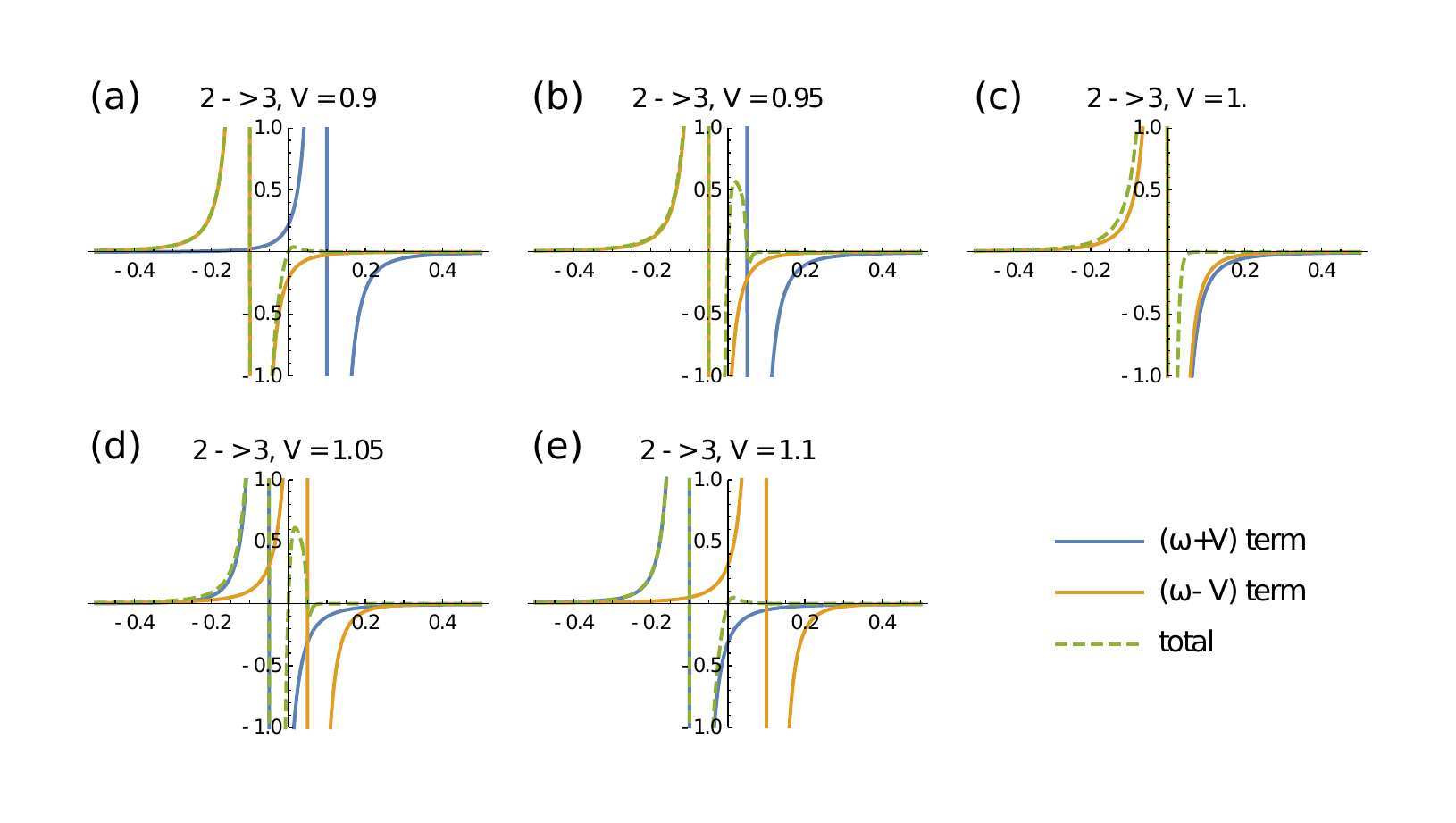}
    \caption{Frequency dependency of the integrand in Eq.~\eqref{I_nnplus1} for varying voltages around the resonant voltage $V=1.0$~a.u. from (a) $V=0.9$ to (e) $V=1.1$. These integrands correspond to the bond current $I_{23}$ for the meta coupling configuration. The two curves, `$\omega+V$' and `$\omega-V$', represent the two terms in the square bracket of the integrand. The `total' curve is the sum of the two terms including the Fermi function. In contrast to Fig.~\ref{fig:integrands}, here we set $\Gamma=0.1$ and $\beta=100$ corresponding exactly to the calculation in the main text [cf.~Fig.~\ref{fig:I_SS}(b)]. We also shrink the frequency axis around $\omega=0$ for better visualization. A competition between the terms arriving from the couplings to source and drain electrodes is clearly visible.}
    \label{fig:interference}
\end{figure}

\begin{figure}[h!]
    \centering
    \includegraphics[width=0.95\textwidth]{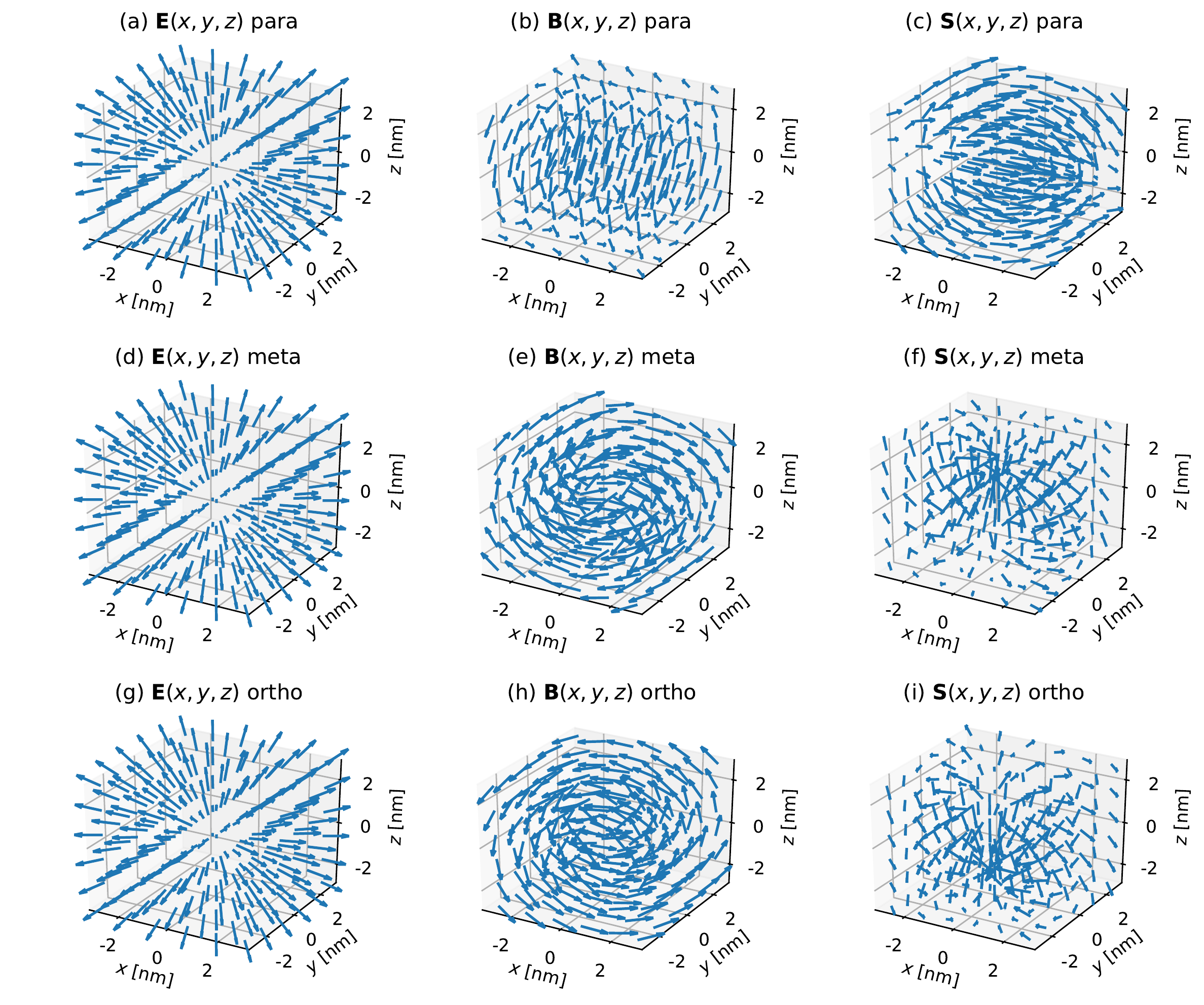}
    \caption{Vector fields $\mathbf{E}$ (left), $\mathbf{B}$ (center), and $\mathbf{S}$ (right) in a cartesian coordinate system up to distances $R=10d$, where $d=2a=2.8$~\AA~is the molecular diameter. The origin is fixed at the center of the benzene molecule (not depicted). The fields correspond to Fig.~\ref{fig:radiationV1} in the main text for para (top row), meta (center row), and ortho (bottom row) coupling configurations. The bias voltage is set to $V=1.0$~a.u. For better visualization of the vector field lines, we have used fixed-length arrows, i.e., they do not correspond to the field strength.}
    \label{fig:vectorfieldsV1}
\end{figure}

\begin{figure}[h!]
    \centering
    \includegraphics[width=0.95\textwidth]{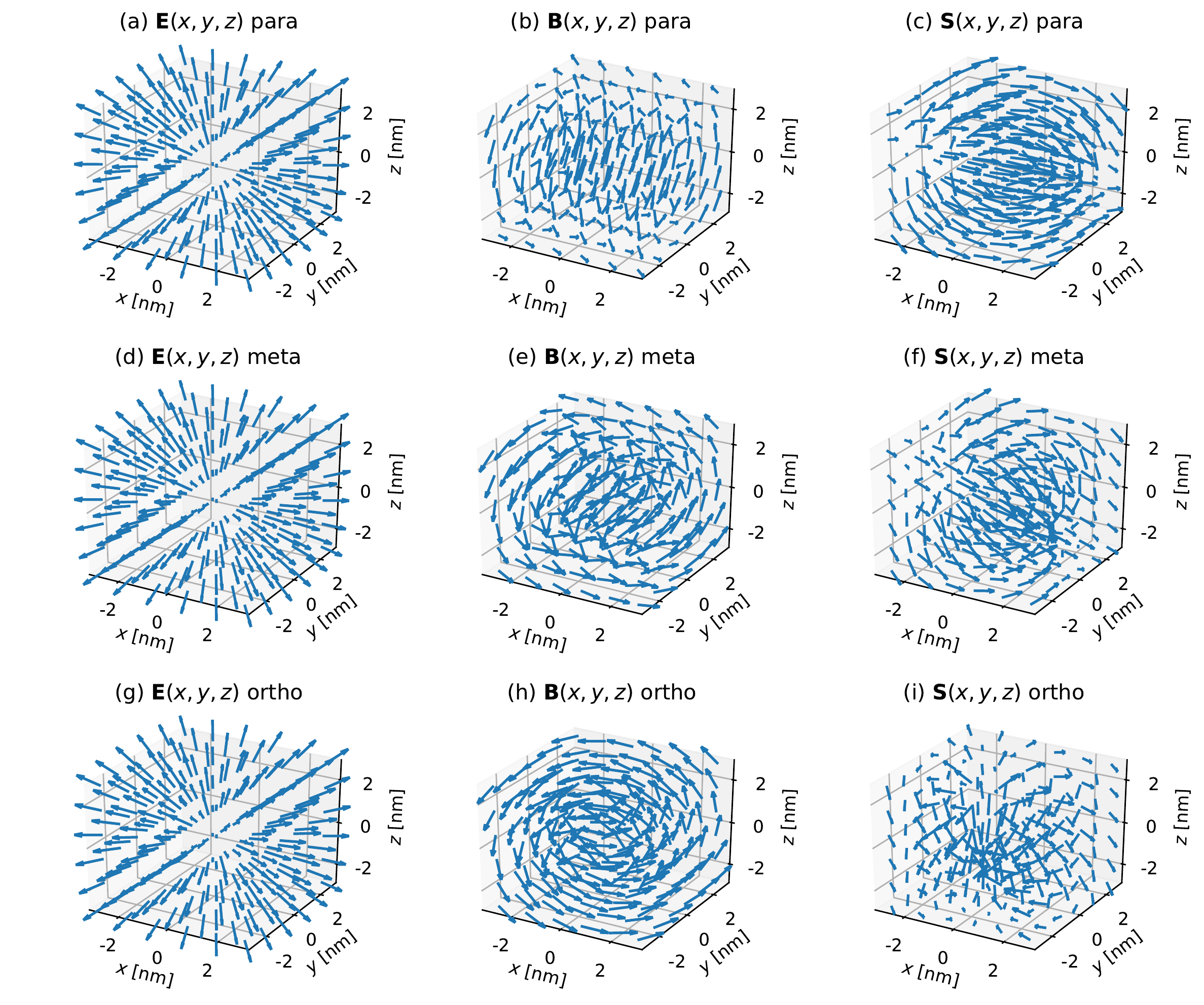}
    \caption{Same as Fig.~\ref{fig:vectorfieldsV1} but with the bias voltage set to $V=2.0$~a.u.}
    \label{fig:vectorfieldsV2}
\end{figure}

\end{document}